\documentclass[prd,superscriptaddress,amsfonts,amssymb,amsmath,showpacs,twocolumn]{revtex4-2}
\usepackage{bm}
\usepackage{amsfonts}
\usepackage{latexsym}
\usepackage[latin1]{inputenc}
\usepackage{graphicx}
\usepackage{amsmath}
\usepackage{palatino}
\usepackage{mathpazo}
\usepackage{textcomp}
\usepackage{float}
\usepackage{booktabs}
\usepackage{dcolumn}
\usepackage{hyperref}
\usepackage{bigints}
\usepackage{subfigure}
\usepackage[english]{babel}
\usepackage[autostyle]{csquotes}
\usepackage{caption}

\hypersetup{colorlinks,citecolor=red}

\usepackage{amsmath}
\usepackage{xcolor}
\usepackage{orcidlink}
\usepackage{commath}

\def\jnl@style{\it}
\def\aaref@jnl#1{{\jnl@style#1}}

\def\aaref@jnl#1{{\jnl@style#1}}

\def\aj{\aaref@jnl{AJ}}                   
\def\apj{\aaref@jnl{ApJ}}                 
\def\apjl{\aaref@jnl{ApJ}}                
\def\apjs{\aaref@jnl{ApJS}}               
\def\apss{\aaref@jnl{Ap\&SS}}             
\def\aap{\aaref@jnl{A\&A}}                
\def\aapr{\aaref@jnl{A\&A~Rev.}}          
\def\aaps{\aaref@jnl{A\&AS}}              
\def\mnras{\aaref@jnl{Mon.~Not.~Roy.~Astron.~Soc.}}             
\def\prd{\aaref@jnl{Phys.~Rev.~D}}        
\def\prc{\aaref@jnl{Phys.~Rev.~C}}  
\def\prl{\aaref@jnl{Phys.~Rev.~Lett.}}    
\def\qjras{\aaref@jnl{QJRAS}}             
\def\skytel{\aaref@jnl{S\&T}}             
\def\ssr{\aaref@jnl{Space~Sci.~Rev.}}     
\def\zap{\aaref@jnl{ZAp}}                 
\def\nat{\aaref@jnl{Nature}}              
\def\aplett{\aaref@jnl{Astrophys.~Lett.}} 
\def\apspr{\aaref@jnl{Astrophys.~Space~Phys.~Res.}} 
\def\physrep{\aaref@jnl{Phys.~Rep.}}      
\def\physscr{\aaref@jnl{Phys.~Scr}}       
\def\commat{\aaref@jnl{Comm.~Math.~Phys.}}              
\def\science{\aaref@jnl{Science}}               
\def\cqg{\aaref@jnl{Classical Quant.~Grav.}}            
\def\jpcs{\aaref@jnl{JPCS}}                                     
\def\ijmpd{\aaref@jnl{Int.~J.~Mod.~Phys.~D}}                    
\def\grg{\aaref@jnl{Gen.~Relat.~Gravit.}}               
\def\rpp{\aaref@jnl{Rep.~Prog.~Phys.}}          
\def\npa{\aaref@jnl{Nucl.~Phys.~A}}        
\def\lrr{\aaref@jnl{Living Rev.~Rel.}}                   
\def\jcap{\aaref@jnl{J.~Cosmology Astropart.~Phys.}}    
\def\rmp{\aaref@jnl{Rev.~Mod.~Phys.}}   
\def\epjc{\aaref@jnl{Eur.~Phys.~J.~C}}

\allowdisplaybreaks[1]

\begin{document}

\color{black}

\title{Study of charged gravastar model in $f(\mathcal{Q})$ gravity}

\author{Debasmita Mohanty\orcidlink{0009-0006-8118-5327}}
\email{newdebasmita@gmail.com}
\affiliation{Department of Mathematics, Birla Institute of Technology and
Science-Pilani,\\ Hyderabad Campus, Hyderabad-500078, India.}

\author{Sayantan Ghosh\orcidlink{0000-0002-3875-0849}}
\email{sayantanghosh.000@gmail.com}
\affiliation{Department of Mathematics, Birla Institute of Technology and Science-Pilani,\\ Hyderabad Campus, Hyderabad-500078, India.}

\author{P.K. Sahoo\orcidlink{0000-0003-2130-8832}}
\email{pksahoo@hyderabad.bits-pilani.ac.in}
\affiliation{Department of Mathematics, Birla Institute of Technology and
Science-Pilani,\\ Hyderabad Campus, Hyderabad-500078, India.}

\date{\today}

\begin{abstract}

In recent days gravastar has been a very lucrative alternative to black holes, as it does not suffer from the singularity problem as well and it is based on sound physical grounds. Modified Symmetric teleparallel equivalent of gravity also has seen quite a few successes in recent years both in cosmology as well as in astrophysical objects like black holes and wormholes. In this paper, we have considered the charged gravastar in $f(\mathcal{Q})$ formulation and have solved it fully analytically and found various physical characteristics like energy density, entropy and EoS for the gravastar. We have used the Israel junction condition to make some phenomenological predictions regarding the potential of the thin shell around the gravastar. We have also studied the deflection of the angle caused by the gravastar. Finally, we conclude by noting how future radio telescopes could detect the gravastar shadow and how can one distinguish it from the black hole event horizon.


\end{abstract}

\maketitle
\section{Introduction}\label{sec:I}

Gravitational Vacuum Condensate Stars or in short just gravastar is a lucrative alternative to black holes which was first proposed by Mazur and Mottala \cite{Mottola/2002, Mazur/2004} in the early twenty-first century. Black holes are the most fascinating object in general relativity not just because they are the exact solutions of Einstein's equation but also the fact that they have entropy which comes purely from quantum mechanics \cite{Hawking/1974}. It was shown by Roger Penrose \cite{Penrose/1965} by using topological arguments that even a non-spherical symmetric collapse could lead to black hole singularity. Black holes or massive compact objects have been observationally found by Ghez \cite{Ghez/1998} and Gillessen  \cite{Gillessen/2009}. We note that Penrose, Ghez, and Gillessen have shared the 2020 Nobel Prize in physics for their effort. However, there are several problems with taking the final stage of a massive star collapse to be a black hole. First, as shown by Penrose under reasonable energy conditions, the black holes or the collapsing stars would always form a singularity, and in singularity, the metric is not invertible so the assumptions of general relativity itself fail. We also note that Hawking radiation of black hole could give Hawking temperature associated with the black hole as
$T_{BH}=\frac{\hbar c^3}{8\pi G k_B M}$ also the black hole entropy could be shown as $S_{BH}=\frac{4\pi k_B  GM^2}{\hbar c}$. We first note that the immediate problem with such a system in the context of thermodynamics is that the energy is given as $dE=T_{BH}dS_{BH}=\frac{\hbar c^3}{8\pi G k_B M} d(\frac{4\pi k_B  GM^2}{\hbar c}$), we note a crucial observation that the reduced plank constant ($\hbar$) does get cancelled out. So we can claim that the thermodynamic quantities behave as classical (non-quantum). This creates a serious paradoxical problem as we noted by Hawking \cite{Hawking/1976} that the energy (or $M$ in natural units) is inversely proportional to temperature. So this leads to negative heat capacity. This is problematic as we already discussed that even though the origin of black hole thermodynamics laws is quantum mechanical the thermodynamic energies behave as a classical theory, and we know from classical thermodynamics that such negative heat capacity could not make a system equilibrium. This paradox can be naturally explained by the gravastar model as shown in \cite{Mazur/2004}.\\
The idea of gravastar comes from the fact that like in a condensed matter system during the phase transition (when temperature is close to critical temperature ($T\approx T_c$) the entire system goes through a phase transition like ferromagnet to paramagnet etc.) As the quantum mechanics gets non-negligible near the event horizon one can expect that the collapsing dust particles also behave like a quantum many-body system. It is also well known that a system of bosonic particles in very low temperatures goes to Bose-Einstein condensation. The idea of Mazur and Motalla was to use this intuition to explain the dust collapse without the pathological problems encountered by the black hole. We also note that the use of Bose-Einstein condensation in an astrophysical context is not at all new it is widely believed that the core or the neutron star is Bose-Einstein condensate, also there are many studies of exotic stars like Bose stars  \cite{Panotopoulo/2018} which are expected to form due to Bose-Einstein condensation.  \\
We note that the entropy of an ordinary object with the same radius and mass as the black hole the entropy is given by $S=\frac{4k+4}{7k+1}S_{BH}$ as shown in \cite{Zurek/1984} ($k$ is the EoS for the object). Such an expression of entropy is a unique feature of gravastar as it in the surface matches with Hawking's semi-classical treatment of black hole entropy, on the other hand, it is non-constant like other singularity-free astrophysical objects.  As we note, for gravastar to mimic the entropy formula of a black hole given by Hawking, we have to take $k=1$, so we take the outer shell of the gravastar as a stiff matter where $k=1$.  \\
Modified symmetric teleparallel gravity or $f(\mathcal{Q})$ gravity has been proposed by Jimenez et al. \cite{Jimenez/2018}. It has been successfully used in the cosmological context to explain the late time acceleration \cite{Solanki/2021, Koussour/2022}, also, there is much literature that has studied the observation data using $f(\mathcal{Q})$ gravity \cite{Mandal/2020}. $f(\mathcal{Q})$ gravity has also been very successful in explaining the properties of various astrophysical objects like black holes \cite{Ambrosio/2022} and wormholes \cite{Hassan/2023} etc. We also note that the study of gravastars in modified gravity is not new, there are several articles in the literature that explore the gravastars in various gravity like $f(R,T)$ gravity \cite{Das/2017, Pramit/2021}, $f(T)$ gravity \cite{ Das/2020}, $f(\mathcal{Q})$ gravity \cite{S/2023}, Rastall-Rainbow gravity \cite{Debnath/2021} etc. The property of charged black holes in modified gravity is of particular interest as the field equation can be most often solved (in rotating or Kerr black holes in modified gravity background typically are not exactly solvable due to the high non-linearity of the field equation). The charged black holes in modified gravity have been also extensively studied by \cite{Javed/2023}. We note that due to the black hole no-hair theorem \cite{Will/2008}, it can be shown that any black hole can have at most three variables those are mass ($M$), charge (${Q}$), and angular momentum ($J$) \cite{WIsrael/1967, Werner/1968}. We note that even though charged gravastar in various modified gravity like $f(T), f(R)$ has already been studied, to the best of our knowledge there has been no study about the charge gravastar in $f(\mathcal{Q})$ gravity. \\
In this paper, we have taken the modified symmetric teleparallel gravity in order to model the gravastar in a charged background. As we know from the black hole uniqueness theorem a black hole has ``no hair" that is a black hole can be fully classified by its charge ($Q$) mass ($M$) and angular momentum ($J$). As studying rotating gravastar is very difficult in fact as the outside is a Karr metric we take a charged gravastar and use Reissner-Nordstrom solutions outside the gravastar to solve the system. We have also used the Israel junction condition to find the potential as well as the mass of the thin shell, in this process we have found the deflection of such a gravastar, giving us the phenomenological tool to distinguish it from black holes. We have also calculated the ADM mass for the inner and outer regions of the gravastar in order to incorporate the ``effective mass" for gravitational interaction.\\
Gravastar is made up of three regions i.e. interior, thin shell and outer region. We take inner and outer radii of each layer being $r_1$ and $r_2$, respectively, and $r_1< r_2$. The three layers are as follows:

\begin{itemize}
   \item  Interior region: $0\leq r <r_1$ with a de Sitter space-time defined equation of state (EoS), $p = -\rho$.

\item  Shell region: $r_1< r < r_2$ with EoS $p =\rho $, an intermediary thin layer made of an ultra-stiff perfect fluid.

\item Exterior region: $r_2 < r$ with EoS $p =  0$ (vacuum), which is described by the Schwarzschild solution.
\end{itemize}

We would use the Israel junction condition on this thin shell to study the behaviour of the gravastar.

Within the context of Einstein's general relativity, a lot of research on gravastar models has been published in the literature. Gravastar solutions with constant pressures and the equation of state have been discussed by DeBenedictis et al. \cite{DeBenedictis/2006}. In Refs \cite{ Bhar/2014, Ghosh/2017, Rahaman/2015}, the gravastar model in higher-dimensional space-time has been discussed.

The format of our paper is as follows: We provided a quick overview of the gravastar model and the most current research on it in sec \ref{sec:I}. In sec \ref{sec:II}, we shall describe the basic framework of  $f(\mathcal{Q})$ gravity. In sec \ref{sec:III}, Maxwell's equation in curved space-time is provided. The summary of the Einstein-Maxwell field equation in $f(\mathcal{Q})$ gravity is provided in sec \ref{sec:IV}. In sec \ref{sec:V}, we discuss the outline of the conformal killing vectors. In sec \ref{sec:VI}, the field equation solution for different regions with varying EoS is provided. Following that, in sec \ref{sec:VII}, we looked at the junction condition and deflection angle. The physical features of the model are described in sec \ref{sec:VIII}. In sec \ref{sec:IX}, we have discussed the ADM mass. Finally, in sec \ref{sec:X}, we present the conclusions of our analysis.

\section{$f(\mathcal{Q})$ gravity} \label{sec:II}

The theory of $f(\mathcal{Q})$ gravity deals with a metric-affine space-time, where the metric tensor $g_{\mu \nu}$ and connection $\Gamma^\lambda_{\mu \nu}$ are treated as independent entities. In this theory, the non-metricity of the connection is defined by,
\begin{equation}\label{eq:1} 
    \mathcal{Q}_{\alpha \mu \nu}=\bigtriangledown_{\alpha} g_{\mu \nu}=\partial_{\alpha} g_{\mu \nu} -\Gamma^{\lambda}\,_{\alpha \mu} \,g_{\lambda \nu}- \Gamma^{\lambda}\,_{\alpha \nu} g_{\mu \lambda} .
\end{equation}
So, basically, the affine connection can be consisted of the following three independent components,
\begin{equation}\label{eq:2}
\Gamma^{\lambda}\,_{\mu \nu}=  \{ {}^\lambda\,{}_{\mu \nu}\}+ K^{\lambda}\,_{\mu \nu} +L^{\lambda}\, _{\mu \nu}. 
\end{equation}
The symbols $\{ ^{}\lambda{}_{\mu \nu}\}$ and $K^{\lambda}\,_{\mu \nu}$ represent the Levi-Civita connection and contortion, respectively and they are defined as follows, 
\begin{equation} \label{eq:3}
    \{ {}^\lambda\,{}_{\mu \nu}\} \equiv \frac{1}{2} g^{\lambda \beta} \left( \partial_{\mu} g_{\beta \nu}+\partial_{\nu} g_{\beta \mu}-\partial_{\beta} g_{\mu \nu} \right),
\end{equation}

\begin{equation}\label{eq:4}
K^{\lambda}\,_{\mu \nu} \equiv \frac{1}{2} T^{\lambda}\,_{\mu \nu}+ T_(\mu^{\lambda}\,_\nu) ,
\end{equation}   
where $T^{\lambda}\,_{\mu \nu}$ is the antisymmetric part of the affine connection, which is defined as $T^{\lambda}\,_{\mu \nu} \equiv 2 \Gamma^{\lambda}\,_{[\mu \nu]}$. \\
The disformation $L^{\lambda}\,_{\mu \nu}$ is defined by,
\begin{equation}\label{eq:5}
 L^{\lambda}\,_{\mu \nu} \equiv \frac{1}{2} \mathcal{Q}^{\lambda}\,_{\mu \nu}- \mathcal{Q}(_{\mu}\,^{\lambda}\,_{\nu} ) .
\end{equation}

Next, the non-metricity conjugate is defined by,
\begin{multline}\label{eq:6}
    P^{\alpha}\,_{\mu \nu}=-\frac{1}{4} \mathcal{Q}^{\alpha}\,_{\mu \nu}+\frac{1}{2} \mathcal{Q}(_{\mu}\,^{\alpha}\,_{\nu})+\frac{1}{4}(\mathcal{Q}^{\alpha}-\Tilde{\mathcal{Q}}^{\alpha}) g_{\mu \nu} \\ 
    -\frac{1}{4} \delta^{\alpha}(_\mu \mathcal{Q}_{\nu}), 
\end{multline}

where $\mathcal{Q}_{\alpha} \equiv \mathcal{Q}_{\alpha}\,^{\mu}\,_{\mu}$ and  $\Tilde{\mathcal{Q}}_{\alpha} \equiv \mathcal{Q}^{\mu}\,_{\alpha \mu}$ are  the two independent traces of non-metricity tensor.\\
The non-metricity scalar is finally defined as follows:

\begin{equation}\label{eq:7}
    \mathcal{Q}\equiv - \mathcal{Q}_{\alpha \mu \nu} P^{\alpha \mu \nu} .
\end{equation}
Along Lagrange multipliers, $f(\mathcal{Q})$ gravity is given by the following action \cite{Jimenez/2018}:

\begin{multline}\label{eq:8}
    S=\int \sqrt{-g} d^4x \left[\ \frac{1}{2} f(\mathcal{Q}) + \lambda_{\alpha}\,^{\beta \mu \nu} R^{\alpha}\,_{\beta \mu \nu} + \lambda_{\alpha}\,^{\mu \nu} T^{\alpha}\,_{\mu \nu}\right.\\
    \left.+ \mathcal{L}_m + \mathcal{L}_e \right]\, , 
\end{multline}

where $g$ is the determinant of the metric $g_{\mu \nu}$, $f(\mathcal{Q})$ represents the arbitrary function of the non-metricity $\mathcal{Q}$. $ \lambda_{\alpha}\,^{\beta \mu \nu}, \mathcal{L}_m, $  and      $\mathcal{L}_e$ represents the Lagrange multipliers, matter Lagrangian and Lagrangian for the electromagnetic respectively.\\

The field equation is obtained by varying action\eqref{eq:8} with respect to the metric \cite{Wang/2022},

\begin{multline}\label{eq:9}
-T_{\mu \nu} + E_{\mu \nu}=\frac{2}{\sqrt{-g}} \bigtriangledown_{\alpha} \left( \sqrt{-g} f_{\mathcal{Q}} P^{\alpha}\,_{\mu \nu}\right) +\frac{1}{2} g_{\mu \nu} f\\
+f_{\mathcal{Q}} \left(  P_{\mu \alpha \beta} {\mathcal{Q}}_{\nu} \,^{\alpha \beta}  - {2\mathcal{Q}}_{\alpha \beta \mu} P^{\alpha \beta}\,_{\nu}\right), 
\end{multline}

where a subscript appears $\mathcal{Q}$ represents a derivative of $f(\mathcal{Q})$ with respect to $\mathcal{Q}$, namely $f_{\mathcal{Q}} \equiv \partial_{\mathcal{Q}} f(\mathcal{Q})$. $E_{\mu \nu}$ is the energy-momentum tensor of the electromagnetic field. The energy-momentum tensor is defined as follows:

\begin{equation}\label{eq:10}
   T_{\mu \nu} \equiv - \frac{2}{\sqrt{-g}} \frac{\delta(\sqrt{-g} \mathcal{L}_m)}{\delta g ^{\mu \nu}} ,
\end{equation}

We note that $E_{\mu \nu}$, the energy-momentum tensor of the electromagnetic field can be found as \cite{Yousaf/2019},
\begin{equation} \label{eq:11}
    E_{\mu \nu} (EM)= \frac{1}{4 \pi} \left[- F^\zeta_\mu F_{\nu \zeta} +\frac{1}{4} F_{\eta\zeta} F^{\eta\zeta} g_{\mu \nu}\right] 
\end{equation}

Now varying equation \eqref{eq:8} with respect to the connection we obtain,
\begin{equation}\label{eq:12}
\bigtriangledown _{\rho} \lambda _{\alpha}\,^{\nu \mu \rho} + \lambda_{\alpha} \, ^{\mu \nu} = \sqrt{-g} f_{\mathcal{Q}} P^{\alpha}\,_{\mu \nu} +H_{\alpha}\, ^{\mu \nu}, 
\end{equation}
 where $H_{\alpha}\, ^{\mu \nu}$ is the hyper momentum tensor density, which is defined as follows \cite{Wang/2022}:

\begin{equation} \label{eq:13}
   H_{\alpha}\, ^{\mu \nu} = \frac{-1}{2} \frac{\delta (\mathcal{L}_m + \mathcal{L}_e)}{\delta \Gamma^{\alpha}\,_{\mu \nu}}. 
\end{equation}
Using the anti-symmetry property of $\mu$ and $\nu$ in the Lagrangian multiplier coefficients, equation \eqref{eq:12} can be simplified to

\begin{equation}\label{eq:14}
    \bigtriangledown_{\mu} \bigtriangledown_{\nu} \left( \sqrt{-g} f_{\mathcal{Q}} P^{\mu \nu}\,_{\alpha} +H_{\alpha}\, ^{\mu \nu}\right) =0. 
\end{equation}
Taking $    \bigtriangledown_{\mu} \bigtriangledown_{\nu} H_{\alpha}\, ^{\mu \nu}=0 $ (according to Ref.  \cite{Jimenez/2018} ), the above equation becomes,

\begin{equation} \label{eq:15}
    \bigtriangledown_{\mu} \bigtriangledown_{\nu} \left( \sqrt{-g}f_{\mathcal{Q}} P^{\mu \nu}\,_{\alpha} \right)=0. 
\end{equation}
The affine connection has the following form in the absence of curvature and torsion.

\begin{equation} \label{eq:16}
    \Gamma^{\alpha}\,_{\mu \nu} = \left( \frac{\partial x^{\alpha}}{\partial \theta^{\lambda}} \right) \partial_{\mu} \partial_{\nu}\, \theta^{\lambda}. 
\end{equation}
We can choose a special set of coordinates, the so-called coincident gauge, so that $\Gamma^{\alpha}\,_{\mu \nu}=0 $, in this case. Consequently, non-metricity reduces to,

\begin{equation}\label{eq:17}
    {\mathcal{Q}}_{\alpha \mu \nu} = \partial_{\alpha} g_{\mu \nu}. 
\end{equation}
This greatly simplifies the calculation, since only the metric variable is essential. However, the action is no longer diffeomorphism invariant with the exception of STGR \cite{Beltran/2020}. One can avoid the issue by employing the covariant formulation of $f(\mathcal{Q})$ gravity. Since the affine connection in equation \eqref{eq:16} is solely inertial, the covariant formulation could be applied by first determining the affine connection in the absence of gravity \cite{Zhao/2022}. However, as demonstrated in this work, the off-diagonal component of the field equations in the coincident gauge would impose stringent constraints on $f(\mathcal{Q})$ gravity, thereby providing nontrivial functional forms of $f(\mathcal{Q})$.

\section{Maxwell's equation in curved space time} \label{sec:III}

We note that all four of Maxwell's equations can be expressed as,
\begin{equation} \label{eq:18}
    \frac{\partial F_{\mu \nu}}{\partial x_{\nu}}=\mu_0 J_{\mu}.
\end{equation}

\begin{equation} \label{eq:19}
    \frac{\partial F_{\mu \nu}}{\partial x_{\lambda}} +\frac{\partial F_{\nu \lambda}}{\partial x_{\mu}} +\frac{\partial F_{\lambda \mu }}{\partial x_{\nu}} =0.
\end{equation}

For a free electromagnetic field, Lagrangian density is defined as follows,

\begin{equation} \label{eq:20}
  L_{EM} =-\frac{1}{16 \pi} F^{\mu \nu} F_{\mu \nu},
\end{equation}
with
\begin{equation} \label{eq:21}
  F_{\mu \nu}= A_{\nu,\mu}  - A_{\mu , \nu}.
\end{equation}

We note that if we vary the Lagrangian with respect to $A_{\mu}$ we will get Maxwell's equation from the Euler-Lagrange equation.\\
Consequently, by varying the above Lagrangian with respect to the metric, we can get its energy-momentum tensor via,
\begin{equation} \label{eq:22}
  E_{\mu \nu}=2 \frac{\partial \mathcal{L}_m}{\partial g^{\mu \nu}} - \mathcal{L}_m g_{\mu \nu} ,
\end{equation}
So plugging the EM Lagrangian density we get,

\begin{equation*} 
    E_{\mu \nu} (EM)= \frac{1}{4 \pi} \left[- F^{\zeta}_\mu F_{\nu \zeta} +\frac{1}{4} F_{\eta \zeta} F^{\eta \zeta} g_{\mu \nu}\right] 
\end{equation*}

We note that for an isolated static charge, $J^\mu$ is zero. We also note that as we are working with spherical polar coordinate, for the field strength of a static charge only component of interest is the radial potential as,
\begin{equation*}
    A_0=\phi \neq 0 , \,\, A_2=A_3=A_1=0.
\end{equation*}
Hence, out of all the components of $F_{\mu \nu}$ the only nonzero components will be $F_{01}, F_{10}$. \\
So the electromagnetic field strength tensor can be expressed in the following manner:

\begin{equation} \label{eq:23}
    F_{\mu \nu}=\phi_{,1}  \begin{pmatrix}
    0 & -1 & 0 & 0 \\
    1 & 0 & 0 & 0 \\
    0 & 0 & 0 & 0 \\
    0 & 0 & 0 & 0
\end{pmatrix}
\end{equation}
Now from Maxwell's equation, we get,

\begin{equation} \label{eq:24}
\begin{gathered}
    \frac{\partial F_{\mu \nu}}{\partial x_{\nu}}= {\mu}_0 J_{\mu}=0 , \\
    \implies (g^{00} g^{11} F_{01} \sqrt{-g})_{,1} =0, \\
    \implies  (g^{00} g^{11} \phi \sqrt{-g})_{,1} =0,\\
    \implies e^{-\frac{\lambda +\nu}{2}} \phi_{,1} r^2 = Q = constant.
\end{gathered}
\end{equation}

Thus, the energy-momentum tensor of the electromagnetic field is given as,

\begin{equation} \label{eq:25}
    E^{\mu}_{\nu} = \frac{1}{8 \pi} diag (\frac{Q^2}{r^4}) (1,1,-1,-1).
\end{equation}
We note that here we use $Q$ as a dimensionless parameter in order to get the graph. However, the dimension expression for $Q$ is given by $\frac{Q^2G}{4\pi \epsilon_0c^4}$.

\section{$f(\mathcal{Q})$ GRAVITY WITH SPHERICAL SYMMETRY} \label{sec:IV}

Let's move on to discussing the spherically symmetric configurations. The metric follows the standard format in spherical coordinates. 

\begin{equation}\label{eq:26}
 ds^2=  - e^{A(r)} dt^2 + e^{B(r)} dr^2 + r^2 (d \theta^2 + \sin^2 \theta d\phi^2).
\end{equation}
 
And the non-metricity scalar $Q$  in terms of $r$ is,

\begin{equation}\label{eq:27}
    \mathcal{Q}(r)=-\frac{2 e^{-B}}{r} \left(  A' + \frac{1}{r}\right) . 
\end{equation}

The Einstein-Maxwell field equation in $f(\mathcal{Q})$ gravity are given by ,
\begin{widetext}
\begin{equation}\label{eq:28}
    \frac{f}{2}+ \frac{2}{r} e^{-B} f_{\mathcal{Q} \mathcal{Q}} {\mathcal{Q}}' -f_{\mathcal{Q}}\biggl[\mathcal{Q}+\frac{e^{-B}}{r} (A'+B')+\frac{1}{r^2}\biggr] = \rho + 2\pi E^2, 
\end{equation}

\begin{equation}\label{eq:29}
    -\frac{f}{2}+f_{\mathcal{Q}}\left(  \mathcal{Q}+\frac{1}{r^2}\right)= p_r - 2 \pi E^2, 
\end{equation}
\begin{equation}\label{eq:30}
    -\frac{f}{2}- e^{-B} \left(  \frac{1}{r} +\frac{A'}{2}\right) f_{\mathcal{Q} \mathcal{Q}} {\mathcal{Q}}' + f_{\mathcal{Q}}\biggl[\frac{\mathcal{Q}}{2}-e^{-B} \left({\frac{A''}{2}+\left( \frac{A'}{4} +\frac{1}{2r}\right) \left( A'-B'\right)} \right)\biggr] =p_t + 2 \pi E^2, 
\end{equation}
and
\begin{equation}\label{eq:31}
  \frac{\cot{\theta}}{2}  f_{\mathcal{Q} \mathcal{Q}} {\mathcal{Q}}'=0. 
\end{equation}
\end{widetext}
\begin{equation}\label{eq:32}
\left( r^2 E\right)'= 4 \pi r^2 \sigma(r) e^{\frac{B}{2}}, 
\end{equation}
Thus, the invariant energy is,
\begin{equation}\label{eq:33}
    E(r)=\frac{1}{r^2} \int{4 \pi r^2  \sigma(r) e^{\frac{B}{2}} dr}. 
\end{equation}

Terms $\rho$, $p$, and $E(r)$ denote the matter density, isotropic pressure, and electric field of a charged fluid sphere, respectively. Also, $\sigma(r)$ denotes the surface energy density of thin shell. The prime notation indicates differentiation with respect to the radial coordinate, $r$.

\section{CONFORMAL KILLING EQUATION} \label{sec:V}

The use of conformal symmetry under conformal killing vectors (CKV), represented by the following equation, is an approach that links geometry with matter systematically \cite{Bhar/2021}.

\begin{equation}
    L_{\xi} g_{ik}=\psi g_{ik}, \label{eq:34}
\end{equation}

 where $L$ stands for the Lie derivative operator, and $\psi$ stands for the conformal factor. The metric $g$ is conformally translated onto itself along $\xi$ as a result of the vector $\xi$ producing the conformal symmetry.

 Eq.\eqref{eq:34} yields the killing vector for $\psi = 0 $, a homothetic vector for $\psi$ = constant, and conformal vectors for $\psi$ = $\psi(x, t)$. The underlying space-time is asymptotically flat for $\psi = 0$, which means that the Weyl tensor will also disappear. The conformal killing vectors can therefore be studied to gain a better understanding of space-time geometry.\\
So equation \eqref{eq:34} becomes,
\begin{equation}\label{eq:35}
    L_{\xi} g_{ik}= \xi_{i;k}+ \xi_{k;i}=\psi g_{ik}. 
\end{equation}

The conformal killing equations for line element \eqref{eq:26} are stated as,

    \begin{equation}\label{eq:36}
        A' {\xi}' = \psi ,\,\, B' {\xi}'+ 2 {\xi}',_{1}=\psi, \,\, {\xi}' =\frac{\psi r}{2}, \,\, {\xi}^0=c_1 . 
    \end{equation}

Here the terms `prime' and `comma' refer to the derivative and partial derivative with respect to `$r$' respectively, and $c_1$ is a constant.

The results of the aforementioned equations are,

    \begin{equation}\label{eq:37}
e^A = {c_2}^2 r^2, 
    \end{equation}
    \begin{equation}\label{eq:38}
        e^B = \left( \frac{c_3}{\psi} \right)^2, 
    \end{equation}
    \begin{equation} \label{eq:39}
        {\xi}^i= c_1 \delta^i_0 + \left(  \frac{\psi r}{2}\right) \delta^i_1 .
    \end{equation}

Here $c_2$ and $c_3$ are constants of integration. We note that $\delta^i_j$ denotes the Kronecker delta, that is $\delta^i_j=1$ if $i=j$ else it is $0$.

It is clear  that in order to satisfy the off-diagonal terms in the field equation \eqref{eq:31} we can only take $f(\mathcal{Q})$ of the form,
\begin{equation} \label{eq:40}
    f(\mathcal{Q})= a{\mathcal{Q}}+b ,
\end{equation}
where $a$ and $b$ are constants. We note that at $a=-1$ and $b=0$, it reduces to Einstein's GR.  \\
Substituting equation (\ref{eq:40}) in equations (\ref{eq:28})-(\ref{eq:30}) we get the following equations given below.

\begin{equation*}
 \frac{-a {\psi}^2}{r^2 {c_3}^2} -\frac{2 a \psi {\psi}'}{r {c_3}^2}+\frac{b}{2}+\frac{a}{r^2}=\rho + 2 \pi E^2
\end{equation*}
\begin{equation*}
    \frac{3a {\psi}^2}{r^2{c_3}^2}-\frac{b}{2}-\frac{a}{r^2}=p_r -2\pi E^2
\end{equation*}    

\begin{equation*}
    \frac{a {\psi}^2}{r^2 {c_3}^2}+\frac{2 a \psi {\psi}'}{r{c_3}^2}-\frac{b}{2}=p_t + 2 \pi E^2
\end{equation*}

From the above equations, we get the following Einstein-Maxwell field equations (following the assumptions given in \cite{Ghosh/2017,Yousaf/2019}),

    \begin{equation}  \label{eq:41}
        -\frac{3a \psi {\psi}'}{r {c_3}^2} +\frac{1}{2}  \left( \frac{a}{r^2}+b \right)  = \rho ,
    \end{equation}
\begin{equation}\label{eq:42}
    \frac{2a {\psi}^2}{r^2 {c_3}^2}+ \frac{a \psi {\psi}'}{r {c_3}^2} - \frac{1}{2} \left(  \frac{a}{r^2} +b\right)  = p, 
\end{equation}
\begin{equation}\label{eq:43}
\frac{1}{\pi} \biggl[   \frac{-a {\psi}^2}{2 r^2 {c_3}^2} + \frac{a \psi {\psi}'}{2 r {c_3}^2} +\frac{a}{4 r^2}\biggr] =E^2 . 
\end{equation}

In the next section, we will go over each of the three charged gravastar regions where we solve the aforementioned field equations \eqref{eq:41} - \eqref{eq:43} individually.

\section{Geometry of gravastar} \label{sec:VI}

\subsection{Interior region of the Charged Gravastar} \label{subsec:A}

Let's consider the equation of state (EoS) between the matter density $\rho$ and isotropic pressure $p$.
\begin{equation}\label{eq:44}
  p = -\rho , 
\end{equation}
  as proposed by Mazur and Mottola \cite{Mottola/2002,Mazur/2004}, the metric coefficient for $e^{B}$ can be determined.

By using equation \eqref{eq:44}, we can obtain a differential equation that is linear in the conformal factor $\psi$ as follows, from equations \eqref{eq:41} and \eqref{eq:42}.

\begin{equation}\label{eq:45}
    \frac{a \psi}{r^2 {c_3}^2} \left(  \psi - {\psi}' r \right) =0 . 
\end{equation}

There are two possible values for $\psi$ based on equation \eqref{eq:45}, $\psi=0$ or $\psi=\psi_0 r$,
where $\psi_0$ represents the constant of integration.
We compute the other physical parameters in the charged gravastar model's interior area by using $\psi = \psi_0 r$ since $\psi = 0$ implies asymptotic flatness.

After obtaining the conformal factor $\psi$ expression, we may write the metric coefficients expression in terms of the conformal factor as follows:

\begin{equation}\label{eq:46}
    e^{-B(r)}= {\Tilde{\psi_0}}^2 r^2 ,
    \end{equation}
    \begin{equation}\label{eq:47}
         e^{A(r)} =c_2^2 r^2, 
    \end{equation}
   
where $\Tilde{\psi_0} = \frac{\psi_0}{c_3}$,   and an intriguing feature of this situation is that $e^A$ is directly proportional to $r^2$  but $e^{B}$ is inversely proportional to $r^2$.
One may find the isotropic pressure and the expression for matter density as,

\begin{equation}\label{eq:48}
    \rho= -3 a {\Tilde{\psi_0}}^2 + \frac{1}{2} \left(\frac{a}{r^2}+b \right) =-p.  
\end{equation}
From equation \eqref{eq:48}, we know that ${\Tilde{\psi_0}}^2 < \frac{1}{6 a} \left( \frac{a}{r^2}+b\right) $ since the energy density is positive in the interior region of the gravastar which provides an upper bound for $\Tilde{\psi_0}$. As we go closer to the gravastar's centre, the pressure and density both suffer from central singularities, which is a normal behaviour of the CKV model. Both variables are inversely proportional to $r^2 $ and blow up without bounds.

Again by using $\psi={\psi}_0 r$ in equation (\ref{eq:43}), we get the expression for the electric field as,
\begin{equation}\label{eq:49}
    E^2= \frac{1}{\pi} \left( \frac{a}{4r^2} \right). 
\end{equation}
Using equation (\ref{eq:49}) in equation (\ref{eq:32}) we derived charge density as
\begin{equation}\label{eq:50}
    \sigma(r)= \frac{\sqrt{a} \Tilde{\psi_0}}{8 {\pi}^\frac{3}{2} r}. 
\end{equation}

Here both charge $E$ and charged density $\sigma(r)$ are inversely proportional to $r^2$, while charge does not depend on $ \Tilde{\psi_0}$  but charged density depends on  $\Tilde{\psi_0}$. 
 The active gravitational mass $M(r)$ can be obtained as the following formula,
 
\begin{equation}\label{eq:51}
    M(r)= \pi \left[ \frac{2 b r^3}{3} +4ar(1-\Tilde{\psi_0}^2 r^2) \right]. 
\end{equation}
As  $r$ approaches zero, the mass function $M(r)$ also approaches zero. This means the central singularity does not affect the mass function $M(r)$. We obtain the lower bound for the active gravitational mass, $M(r) > \frac{ 10 \pi a r}{3}$, using the bound for $\Tilde{\psi_0}$.

\subsection{Thin shell of the charged gravastar}\label{subsec:B}

In this region, we assume that the thin shell contains a stiff perfect fluid that obeys the EoS

\begin{equation}\label{eq:52}
p=\rho. 
\end{equation}

This EoS is a particular case of a barotropic EoS with $p = \omega\rho$ and $\omega = 1$. Barotropic fluids are those in which the pressure only depends on the density, i.e., $p = p(\rho)$, and vice versa. They are thought to be implausible, but their simplicity has an educational advantage in showing the many methods employed to solve various systems and ``physically" fascinating problems. In this regard, we would like to point out that Zel'dovich \cite{Zeldovich/1972} first proposed the concept of this type of fluid in relation to a cold baryonic cosmos and it was described as stiff fluid. The spherical collapse of an over-density of a barotropic fluid with a linear equation of state in a cosmic background was examined by Staelens et al. \cite{Staelens/2021} in their study. By assuming an isotropic pressure with and without the cosmological constant $\Lambda$, Rahaman and his collaborators \cite{Rahaman/2014} were able to create a novel class of exact interior solutions in $(2 + 1)$-dimensional space-time using the barotropic EoS. The stiff fluid model has been used earlier by several researchers in the field of astrophysics as well as cosmology \cite{Carr/1975, Braje/2002, Madsen/1992, Ferrari/2007}.

We can obtain the following ODE by using equation \eqref{eq:52} together with equations \eqref{eq:41} and \eqref{eq:42}.

\begin{equation} \label{eq:53}
    \frac{2r \psi {\psi}'}{c_3^2}+ \frac{{\psi}^2}{{c_3}^2}= \frac{1}{2} \left( 1+ \frac{br^2}{a}\right).
\end{equation}

The aforementioned ODE is a linear equation in $\psi^2$, and when integrated, it yields
\begin{equation} \label{eq:54}
\begin{gathered}
 \psi(r)=\frac{\sqrt{3 a r {c_3}^2 +6 a c_4+b r^3 {c_3}^2 }}{\sqrt{6} \sqrt{a} \sqrt{r}},  \\
 \implies \frac{{\psi}^2 [r]}{c_3^2}=\frac{1}{2} +\frac{b}{6 a} r^2 -\frac{\chi}{r}
 \end{gathered} 
\end{equation}

where $\chi=-\frac{c_4}{c_3^2}$ is a constant of integration.

Utilizing the equation \eqref{eq:54} in equations \eqref{eq:37} and \eqref{eq:38} we achieve the metric potential as

 \begin{equation}\label{eq:55}
     e^{-B(r)}=\frac{1}{2}+\frac{b r^2}{6a}-\frac{\chi}{r}, 
 \end{equation} 

\begin{equation}\label{eq:56}
    e^{A(r)}=c_2^2 r^2, 
\end{equation}
 
\begin{figure}[h] 
    \centering
    \includegraphics[scale=0.6]{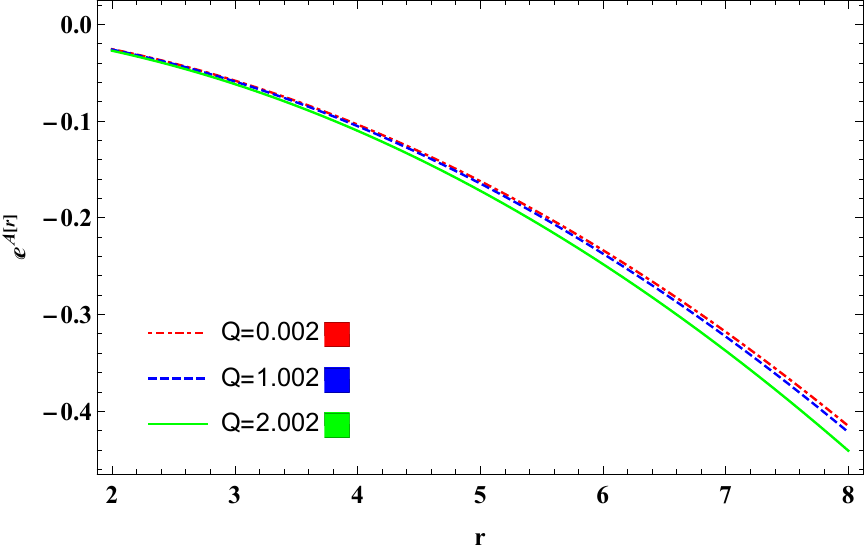}
    \caption{Metric coefficient $e^A$ inside the thin shell}
    \label{fig-1}
\end{figure}

\begin{figure} 
    \centering
    \includegraphics[scale=0.6]{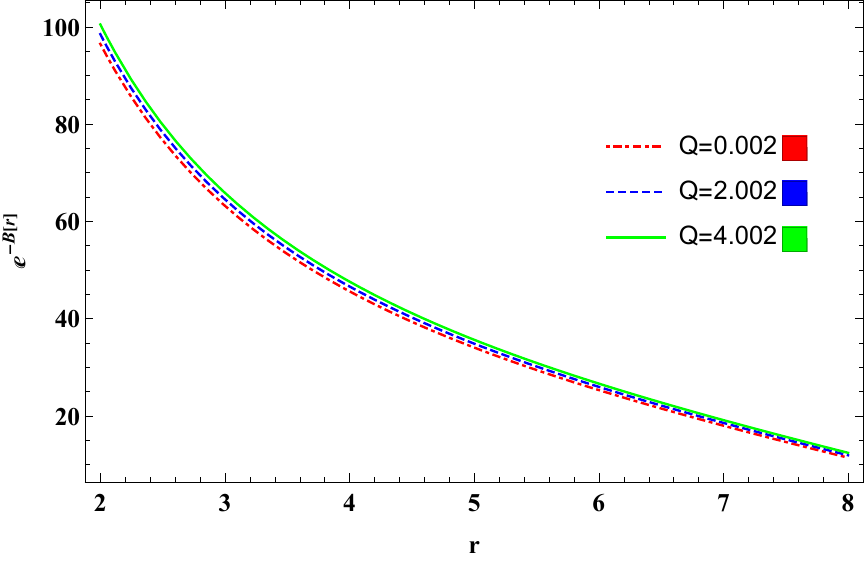}
    \caption{Metric coefficient $e^{-B(r)}$ inside the thin shell}
    \label{fig-2}
\end{figure}

Figures \eqref{fig-1} and \eqref{fig-2} show the profiles of the metric coefficients inside the thin shell.\\
Furthermore, by plugging the EoS $p=\rho$ and using equation \eqref{eq:53} into
the  equations \eqref{eq:41} and \eqref{eq:42} we have obtained the expressions  for matter density and isotropic pressure inside the thin shell as,

\begin{equation}\label{eq:57}
    \rho= \frac{a}{2 r^2}-\frac{3a \chi}{2 r^3}=p . 
\end{equation}

The electric field and charge density of the thin shell are derived as,
\begin{equation}\label{eq:58}
    E^2= \frac{3 a \chi}{4 \pi r^3} , 
\end{equation}
and
\begin{equation}\label{eq:59}
    \sigma(r)= \frac{1}{16 \pi r^2} \sqrt{\frac{3 a \chi}{\pi r}} \sqrt{\frac{1}{2}+\frac{b r^2}{6 a}-\frac{\chi}{r}} . 
\end{equation}
Electric field $E$ depends on $\chi$ and is inversely proportional to $r^{\frac{3}{2}}$. Similar to the previous instance, the central singularity affects both pressure and density. From equation \eqref{eq:59}, we have $\chi<\frac{(3a+br^2)r}{6a}$. Thus, we get $0<\chi<\frac{(3a+br^2)r}{6 a}$ from the above two cases.



\subsection{Exterior region of charged gravastar} \label{subsec:C}

It is assumed that the charged gravastar's exterior obeys the EoS $p=0$, proving that the exterior of the shell is totally vacuum-sealed. The exterior space-time is described by the Reissner-Nordstrom line element given by,
\begin{widetext}
\begin{equation}\label{eq:60}
ds^2=-\left( 1-\frac{2M}{r}+\frac{Q^2}{r^2}\right) dt^2 +\frac{dr^2}{ \left(1-\frac{2M}{r} +\frac{Q^2}{r^2}\right)} +r^2(d \theta^2 + \sin^2 \theta d\phi^2), 
\end{equation}
\end{widetext}
where $M$ and $Q$ represent the mass and charge of the gravastar, respectively.

\subsection{Boundary Condition} \label{subsec:D}

In the gravastar configuration, there are two junctions or interfaces. Let's call the junction between the intermediate thin shell (at $r=r_1$) and interior space-time junction-$I$ and the junction between the intermediate thin shell and external space-time (at $r=r_2$) junction-$II$. For any stable arrangement, the metric functions at these interfaces must be continuous. To determine the values of the unknown constants of our current study, such as $ c_2$, $\Tilde{\psi_0}$, and $\chi$, we matched the metric functions at these borders.

    \begin{itemize}
        \item \textbf{Junction-I:}
\begin{equation} \label{eq:61}
  {\Tilde{\psi_0}}^2 r_1^2=  \frac{1}{2}+\frac{b r_1^2}{6a}-\frac{\chi}{r_1}.
\end{equation}

    \item \textbf{Junction-II:}
    \begin{equation} \label{eq:62}
        c_2^2 r_2^2=-\left( 1-\frac{2M}{r_2}+\frac{Q^2}{r_2^2} \right), 
    \end{equation}
    \begin{equation}\label{eq:63}
    \frac{1}{2}+\frac{b r_2^2}{6a}-\frac{\chi}{r_2}=-\left( 1-\frac{2M}{r_2}+\frac{Q^2}{r_2^2} \right).  
    \end{equation}
     \item \textbf{Obtained Constants :}

\begin{equation} \label{eq:64}
    c_2= \frac{1}{r_2} \sqrt{-1+\frac{2M}{r_2}-\frac{Q^2}{r_2^2}}.
\end{equation}

\begin{equation} \label{eq:65}
    \chi=\frac{3r_2}{2}-2M+\frac{Q^2}{r_2}+\frac{b r_2^3}{6 a}.
\end{equation}

\begin{equation}\label{eq:66}
\Tilde{\psi_0}=\frac{1}{r_1} \sqrt{ \left( \frac{1}{2}+\frac{b r_1^2}{6 a}-\frac{\chi}{r_1} \right)} . 
\end{equation}

    \end{itemize}

We use the mass of the gravastar, $M = 6 M_{\odot}$, inner and outer radii, $r_1=10 km $, $r_2=10.009 km$, and $Q = 0.002$ (in natural units) to calculate the values of these constants. Table-I lists the numerical values of $c_2, \chi$, and $\Tilde{\psi_0}$ for various values of model parameter $a$ and $b$ .\\

Because of the surface tensions at the time-like interfaces $r_1$ and $r_2$, the extremely cold radiation fluid in the shell is constrained to the area-$II$ \cite{Mottola/2002, Mazur/2004}. It is important to keep in mind that the hollow sphere with inner and outer radii of $r_1$ and $r_2 = r_1 +\epsilon$ being considered here has $\epsilon << r_1 $, and according to Mazur and Mottola  \cite{Mottola/2002, Mazur/2004}, $\epsilon$ does not exceed the Planck length when we are matching our interior space-time to the exterior Reissner-Nordstrom space-time. Our inside region and the external line element are matched at the boundary.  Although it is clear that the metric coefficients are continuous at $r = a$, this does not imply that the derivatives of these coefficients are similarly continuous at the junction surface.

\begin{table}[ht]
\textbf{\caption{Different numerical values of constants  assuming $r_1=10 \,$ km and $r_2=10.009 $\, km,\, $Q=0.002$.}}
  \label{tab}
\begin{tabular*}{\columnwidth}{@{\extracolsep{\fill}}rccccccr}
\toprule
$a$  & $b$ & $c_2$ & $\chi$ & $\Tilde{\psi_0}$ \\
 \midrule
  $-4.2$ & $3.2$ & $0.0445604$& $0.0445604$ &$0.0482666$\\
  $-3.2$ & $2.2$ & $0.0445604$& $-111.879$ &$0.0479182$\\
  $-2.5$ & $2.3$ & $ 0.0445604$& $-150.734$ &$0.0489987$\\
  $-1.2$ & $1.2$ & $0.0445604$& $-164.104$ &$0.049365$\\
  $-0.5$ & $1.5$ & $0.0445604$& $-498.338$ &$0.057773$\\
  \hline

\end{tabular*}

\end{table}

\section{Junction Condition}\label{sec:VII}
\subsection{Thin shell around gravastar} \label{subsec:A}

Even though we have already discussed that the reason to take the EoS of the thin-shell to be 1 is to match with the black hole entropy formula. We must note that the thin shell has a nonzero dimension which we have also calculated. However, in order to get the potential across the thin shell we note that we have to use the Israel junction condition, which is only valid if the shell is really thin (mathematically 1 less dimension hypersurface has been embedded on the manifold). We also note that the following assumption is justified as beyond shell when we go to de Sitter space inside we expect this to happen via a quantum phase transition mechanism so the shell is barely a junction between two phases it can not be "thick" otherwise it would also be affected during the phase transition but it definitely exists for the reasons explained in the introduction. \\
Study of Junction condition was first done by Sen \cite{Sen/1924}. Later Lanczos \cite{Lanczos/1924}  made several comments about the problem with Den's approach and how to modify it. It was Darmois \cite{Darmois/1927} who first pointed out the exact junction condition but in a rather convoluted way. In this paper, we follow Israel's prescription \cite{Israel/1966, Israel/1967}  which is far more simple and based on sound physical grounds.\\
 As mentioned earlier we use the Israel junction condition to obtain the solutions because there are two different metrics across the thin shell that must match the condition along the boundary (generally, we use the junction condition because the metric along hypersurfaces must be continuous as well as differentiable; as a result, we check both the Christoffel symbol and Riemann curvature tensor to see what the boundary condition leads to, because in thin shell we have to use the discontinuity across the junction to find the potential across the thin shell.)\\

We also observe that $\sum$ would stand for the thin shell or three-manifold. Reissner-Nordstrom solutions outside of the thin shell are designated by $\vartheta^+$, and inside, there is a $\vartheta^-$, and the whole space-time would be $\vartheta^+  \cup \sum \cup \vartheta^-$. We can also concentrate on the surface energy density $\varsigma$ and surface pressure $P$.This can be done as follows:

 We have the interior solutions
\begin{equation*}
 ds^2=  - e^{A(r)} dt^2 + e^{B(r)} dr^2 + r^2 (d \theta^2 + \sin^2 \theta d\phi^2).
\end{equation*}
For  the  exterior  solutions,  we  take  the 
Reissner-Nordstrom solution of the form
\begin{multline*}
 ds^2=-\left( 1-\frac{2M}{r}+\frac{Q^2}{r^2}\right) dt^2 +\frac{dr^2}{ \left(1-\frac{2M}{r} +\frac{Q^2}{r^2}\right)} \\+r^2(d \theta^2 + \sin^2 \theta d\phi^2) .
\end{multline*}

Here, we note that because of Birkhoff's theorem outside metric of a static spherically symmetric gravastar is always the Reissner-Nordstrom metric.
We also note that in both cases, the space-like  component is spherically symmetric, and on the boundary, we can
obtain the FLRW metric
\begin{equation} \label{eq:67}
    ds^2=-d\tau^2 + \textbf{a}(t) d\Omega^2.
\end{equation}
Now, if we use the first junction condition's formula, we obtain
\begin{equation}\label{eq:68}
     K^{\pm}_{ij}= -n^{\pm}_{\nu}\left(\frac{\partial^2 x^{\nu}}{\partial \phi^{i} \partial \phi^{j}}+ \Gamma^l_{km} \frac{\partial x^l}{\partial \phi^i} \frac{\partial x^m}{\partial \phi^j} \right), 
\end{equation}

where $n^{\pm}$ stands for the two-sided unit normal to the surface and $\phi$ stands for the intrinsic co-ordinate in the shell area, which may be represented as,

\begin{equation}\label{eq:69}
    n^{\pm}=\pm \left|g^{lm} \frac{\partial f}{\partial x^{l}} \frac{\partial f}{\partial x^{m}} \right|^{-1/2} \frac{\partial f}{\partial x^{\nu}}, 
\end{equation}

with  $n^{\gamma} n_{\gamma} =1$.\\

We also note that for the second Israel junction condition we can split the $T_{\mu\nu}$ as $T_{\mu\nu}=\Theta(l)T_{\mu \nu}^++\Theta(-l)T_{\mu \nu}^- +\delta(l)S_{\mu\nu}$ \cite{Poisson/2007}, where $S_{\mu\nu}$ denotes the stress-energy tensor for the shell.\\

Now, to evaluate the surface stress and pressure for the thin shell to sustain itself, we use the Lanczos equation


\begin{equation}\label{eq:72}
    S_{ij}=-\frac{1}{8 \pi}(k_{ij}-\delta_{ij} k_{\gamma \gamma}) .
\end{equation}

\begin{figure}[h]
     \includegraphics[scale=0.6]{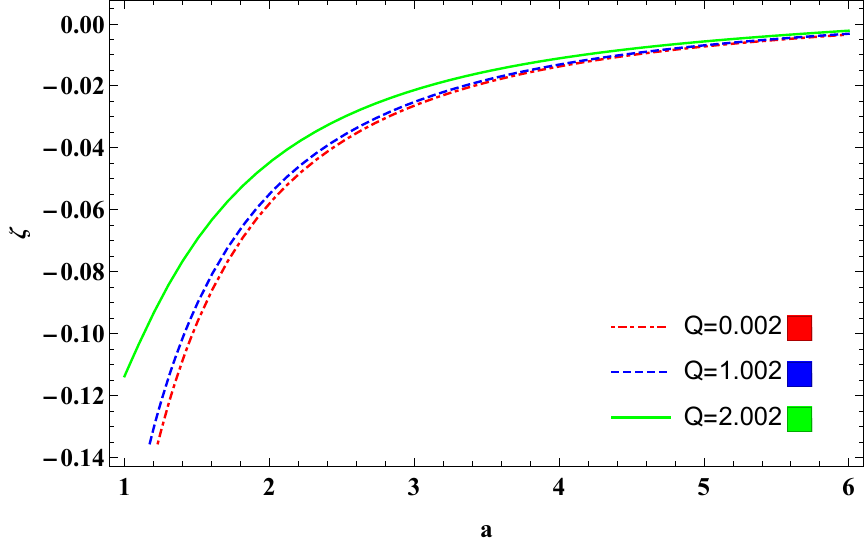}
    \caption{Variation of surface energy density inside the thin shell}
    \label{fig-3}
\end{figure}

\begin{figure}[h]
   \includegraphics[scale=0.6]{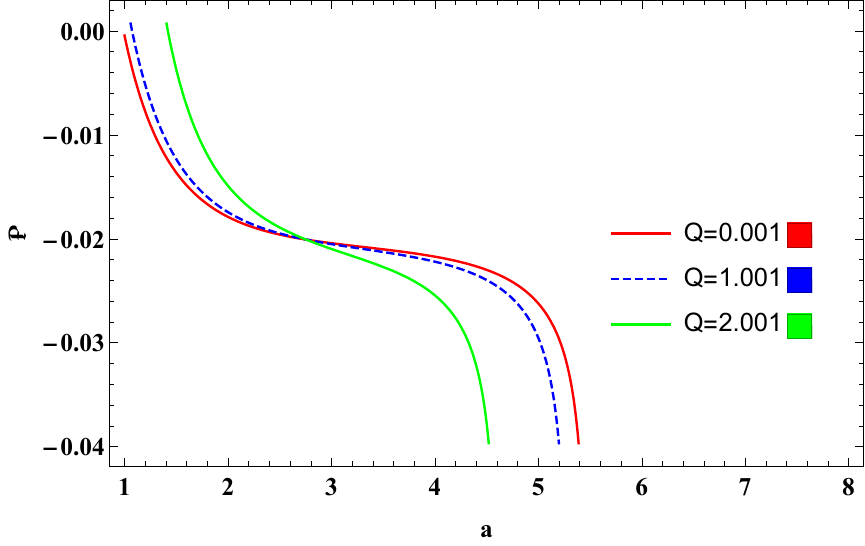}
    \caption{Variation of surface pressure inside the thin shell}
   \label{fig-4}
\end{figure}

Where $i, j =0,2,3$, since at the shell, $r$ is constant. The surface energy tensor can be expressed as $S_{ij}=diag(-\varsigma, P)$. So, the surface energy density $\varsigma$ and pressure $P$ at the junction surface $r=\textbf{a}$ can be obtained by the following equations:




\begin{equation} \label{eq:71}
     \varsigma=-\frac{1}{4 \pi \textbf{a} }\left[\sqrt {f}\right]^{+}_{-}, 
\end{equation}
and
\begin{equation}\label{eq:72}
         P=-\frac{\varsigma}{2}+\frac{1}{16 \pi}\left[\frac{f^{\prime}}{\sqrt {f}}\right]^{+}_{-} .
\end{equation}
Based on the equations (\ref{eq:71}) and (\ref{eq:72}), we can derive the expressions for the aforementioned quantities,

\begin{equation} \label{eq:73}
  \varsigma=  \left( -\frac{1}{{4 \pi  \textbf{a}}} \right)\left(\sqrt{-1+\frac{2 M}{\textbf{a}}-\frac{Q^2}{\textbf{a}^2}}-{\textbf{a}\Tilde{ \psi_0} }\right),
\end{equation}
\begin{equation} \label{eq:74}
 P=\frac{1}{8 \pi \textbf{a}} \left( \frac{-1+\frac{M}{\textbf{a}}}{\sqrt{-1+\frac{2M}{\textbf{a}}-\frac{Q^2}{\textbf{a}^2}}}-2 \textbf{a} \Tilde{\psi_0}\right).
\end{equation}
Figures \eqref{fig-3} and \eqref{fig-4} respectively show the profiles of surface energy density and surface pressure inside the thin shell.

The mass of the thin shell can be obtained using the equation for the surface energy density given by,
\begin{equation} \label{eq:75}
m_{shell}= 4 \pi \textbf{a}^2 \varsigma=  \textbf{a}^2\Tilde{\psi_0}-\textbf{a}\sqrt{-1+\frac{2 M}{\textbf{a}}-\frac{Q^2}{\textbf{a}^2}} .
\end{equation}

As a result, the total mass of charged gravastar in terms of $m_{shell}$ takes the form
\begin{equation}\label{eq:76}
M=\frac{\textbf{a}}{2} \left( 1
+ \frac{Q^2}{\textbf{a}^2}+\frac{m_{shell}^2}{\textbf{a}^2}+{\Tilde{\psi_0}}^2 \textbf{a}^2 - 2m_{shell} \Tilde{\psi_0}\right). 
\end{equation}
We note that by following the prescription given by \cite{M. Sharif, S. D. Forghani}. We can also go further and calculate the potential $V(r)$ by noting that the energy-momentum has a conservation relation:
\begin{equation}\label{eq:77}
    \frac{d}{d\tau}(\varsigma \phi)=p\frac{d\phi}{d\tau}=0\,,
\end{equation}
where $\phi = 4\pi \textbf{a}^2$. From the conservation equation above one can find 
\begin{equation}\label{eq:78}
    \varsigma '=-\frac{2}{\textbf{a}}(\varsigma+p)\,.
\end{equation}
Following the prescription given in \cite{eric}, we note that the last equation has the form of $\dot{\textbf{a}}^2+V(\textbf{a})$, so from the above equation we can get,
\begin{equation}\label{eq:79}
    V(\textbf{a})=\frac{f(\textbf{a})}{2}+\frac{F(\textbf{a})}{2}-\frac{(f(\textbf{a})-F(\textbf{a}))^2}{64\textbf{a}^2\pi^2\varsigma^2}-4\textbf{a}^2\pi^2\varsigma^2. 
\end{equation}

We also note that in our case $f(r)$ and $F(r)$ are given by the following, 
 $$f(r)=\Tilde{\psi}_o^2r^2$$\,
 and 
 $$F(r)=-\left(1-\frac{2M}{r}+\frac{Q^2}{r^2}\right) $$

We note that outside the thin shell, we can take the Reissner-Nordstrom solution as discussed earlier.

\begin{figure}[h]
    \includegraphics[scale=0.6]{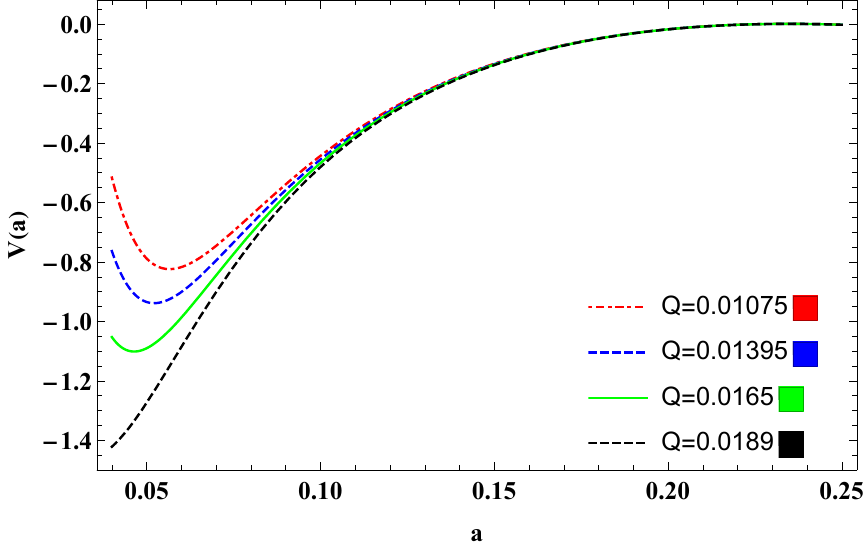}
    \caption{Thin-shell potential across the shell.}
    \label{fig-5}
\end{figure}

We note that the above minimum or the stable circular orbit goes out for higher values of $Q$ we also note that when $Q$ is very small the potential profile is very similar to the Schwarzschild solutions as we expect when $Q=0$ Reissner-Nordstrom solution becomes Schwarzschild solution.


\subsection{Deflection angle for thin shell around gravastar}\label{subsec:B}

It is well known that a massive body acts as a convex lens when light passes around it. As the path of light follows the null geodesics and around massive bodies due to nontrivial metric the null geodesic equation does change.\\
Historically gravitational lensing is the first experimental test of general relativity (1919). Nowadays weak lensing is an essential tool for studying the galaxy cluster.\\
In this paper, we will try to show that if a sufficiently high energy photon created in the interior region can pass through the thin shell (it is physically possible as there is no event horizon), one can use a next-generation Event Horizon like radio telescope to detect such red-shifted photon. We hope in the future one can find the shadow of the inner region of the gravastar just like it has been found in black hole.
For a given metric there are several ways to find the deflection angle. One of them is to use the Hamilton-Jacobi equation for a particle with charge $q$ and mass $m$ given by,
$$g^{ij}\left(\frac{\partial S}{ \partial x^i} + q A_i \right) \left(\frac{\partial S}{ \partial x^k} +q A_k \right)+m^2=0$$
We note that one can integrate the equation for $m=0$, and $q=0$ as we are interested in the photon path. However if one uses the usual null geodesic equation for the following metric,
\begin{equation} \label{eq:80}
 ds^2= -C(r) dt^2 +D(r) dr^2 +F(r) (d\theta^2+ \sin^2 \theta d\phi^2).  
\end{equation}
We use the prescription given by
Virbhadra et al. \cite{Virbhadra/1998}, to calculated the deflection angle as,
\begin{equation} \label{eq:81}
    \alpha(r_0)=I(r_0)-\pi,
\end{equation}

where

\begin{equation} \label{eq:82}
   I(r_0)=  \int^{\infty}_{r_0} \frac{2 \sqrt{D(r)}dr}{\sqrt{F(r)}\sqrt{\frac{F(r) C(r_0)}{F(r_0)C(r)}-1}}. 
\end{equation}
For the interior region,
 \begin{widetext}  
\begin{equation} \label{eq:83}
    C(r)= e^{A(r)}=\frac{r^2}{{r_2}^2}\left( -1+\frac{2M}{r_2}-\frac{Q^2}{{r_2}^2}\right),\\ 
 D(r)= e^{B(r)}= \left( -\frac{Q^2}{r_1^3}+\frac{r^2}{2r_1^2}+\frac{br^2}{6\textbf{a}}-\frac{3 r^2 r_2}{2r_1^3}+\frac{2M r^2}{r_1^3}-\frac{b r^2 r_2^3}{6\textbf{a} r_1^3}\right)^{-1},\,\, F(r)=r^2
 \end{equation}
 \end{widetext}


\begin{widetext}
 Now the above equation  becomes,
    \begin{equation}\label{eq:84}
  I(r_0)_{inside}=2 \bigint \limits^{\infty}_{r_0}  \frac{\left( -\frac{Q^2}{r_1^3}+\frac{r^2}{2r_1^2}+\frac{br^2}{6\textbf{a}}-\frac{3 r^2 r_2}{2r_1^3}+\frac{2M r^2}{r_1^3}-\frac{b r^2 r_2^3}{6\textbf{a} r_1^3}\right)^{-\frac{1}{2}}}{r\sqrt{\left(\frac{r}{r_0}\right)^2 \left[\frac{\left[\left(\frac{r}{r_2}\right)^2(-1+\frac{2M}{r_2}-\frac{Q^2}{r_2^2})\right]_{r=r_0}}{ \left(\frac{r}{r_2}\right)^2\left(-1+\frac{2M}{r_2}-\frac{Q^2}{r_2^2}\right)} \right]-1 }} dr .
   \end{equation}

For the exterior region,
 \begin{equation}\label{eq:85}
     C(r)=-1+\frac{2M}{r}-\frac{Q^2}{r^2},\,\,\,
      D(r)=\left(1-\frac{2M}{r}+\frac{Q^2}{r^2}\right)^{-1},\,\,\,F(r)=r^2.
 \end{equation}

\begin{equation}\label{eq:86}
\begin{gathered}
I(r_0)_{outside}=2 \bigint \limits^{\infty}_{r_0}  \frac{dr}{r \sqrt{1-\frac{2M}{r}+\frac{Q^2}{r^2}}\sqrt{(\frac{r}{r_0})^2 \left(\frac{-1+\frac{2M}{r_0}-\frac{Q^2}{{r_0}^2}}{-1+\frac{2M}{r}-\frac{Q^2}{r^2}} -1\right) }}
       =2 \bigint \limits^{\infty}_{r_0}  \frac{\left(1-\frac{2M}{r}+\frac{Q^2}{r^2}\right)^{-\frac{1}{2}}}{r \sqrt{\left(\frac{r^4}{r_0^4}\right) \left(\frac{-r_0^2+2Mr_0-Q^2}{-r^2+2Mr-Q^2}\right)-1}} dr.
\end{gathered}
\end{equation}
 \end{widetext}

So from the formula of $I(r_0)_{outside}$, we can see that it is exactly similar to the Reissner-Nordstrom black hole. However, the inside is different, as there is no event horizon so if a red-shifted photon can get deflected from inside through the shell, we can expect future radio telescopes can give a resolution between the black hole and gravastar.

\section{Physical features of the model}\label{sec:VIII}

In this section, we will discuss some of the physical characteristics of the constructed structure, including the equation of state, the proper length, the entropy, and the energy levels within the shell's region.  The stiff perfect fluid travels along these space-times through the shell region of the gravastar because the constructed geometry of the gravastar is the matching of two distinct space-times. It will also examine how the electromagnetic field affects several physical characteristics of the charged gravastar in the $f(\mathcal{Q})$ gravity.

\subsection{Proper length of the shell} \label{subsec:A}

According to the theories of Mazur and Mottola \cite{Mottola/2002, Mazur/2004}, the stiff fluid of the shell is situated between the intersection of two space-times. The phase boundary between the inner space and the intermediate thin shell is at $r_1=d$, and the length of the shell extends up to $r_2=d+\epsilon$, which is the phase border between the outer space and the intermediate thin shell. Therefore, the required length or appropriate thickness of the shell as well as the appropriate thickness between these two interfaces can be determined using the following formula:
\begin{equation}\label{eq:87}   
 \begin{gathered} 
     \textit{l}=\int^{d+\epsilon}_{d} \sqrt{e^{B(r)}} dr \\ = \int^{d+\epsilon}_{d} \frac{1}{\sqrt{\frac{1}{2}+\frac{b r^2}{6 a}-\frac{\chi}{r}}} dr= \int^{d+\epsilon}_{d} \frac{1}{f(r)} , 
\end{gathered}
\end{equation}
where $f(r)=\sqrt{\frac{1}{2}+\frac{b r^2}{6 a}-\frac{\chi}{r}}$. \\
One can note that the above formula just restates the $ds^2$ formula for a metric so could be used to find any proper length given the metric.

Evaluating the integral given in equation \eqref{eq:87} poses a significant challenge at present. So to solve the above integral, let us take $\frac{d f(r)}{dr}=\frac{1}{f(r)}$. Hence we get,
\begin{equation}\label{eq:88}
    \textit{l}= f(d+\epsilon)-f(d). 
\end{equation}

\begin{figure}[h]
    \centering
    \includegraphics[scale=0.6]{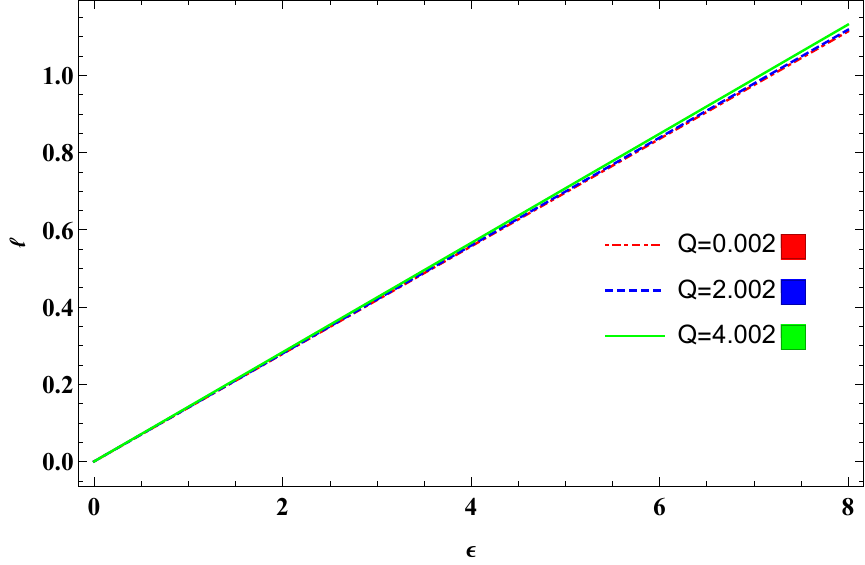}
    \caption{Variation of proper length inside the thin shell}
    \label{fig-6}
\end{figure}

From equation \eqref{eq:88}, by keeping the linear order of $\epsilon$ and expanding $f(d+\epsilon)$ in the Taylor series around `$d$' we get
\begin{equation}\label{eq:89}
    \textit{l}= \epsilon \frac{d f(r)}{dr} \approx  \frac{\epsilon}{\sqrt{\frac{1}{2}+\frac{b r^2}{6 a}-\frac{\chi}{r}}}. 
\end{equation}
The higher-order terms of $\epsilon$ can be disregarded since the value of $\epsilon$ is so small.  Fig-\eqref{fig-6} displays the proper length variation with regard to the thin shell radius.


\subsection{Energy} \label{subsec:B}

The energy of the shell can be calculated by the formula \cite{Bhar/2021},
\begin{equation}\label{eq:90}
    E=\int^{d+\epsilon}_{d} 4 \pi r^2 (\rho+ 2 \pi E^2) dr=2\pi a \epsilon ,\
\end{equation}

\begin{figure}[h]
    \includegraphics[scale=0.6]{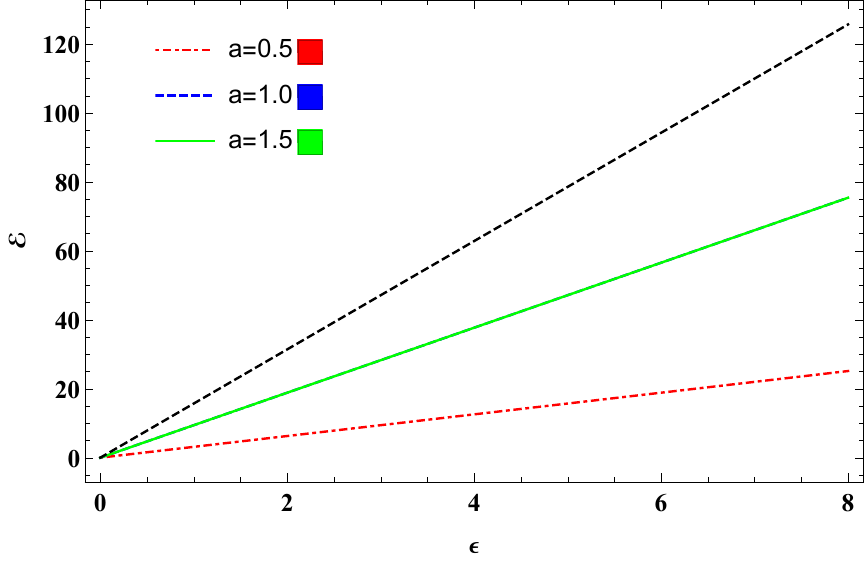}
    \caption{ Variation of  energy  inside the thin shell }
    \label{fig-7}
\end{figure}

In this, the energy and shell thickness are directly correlated. The unit of energy is also ``km" because, as we can see from equation \eqref{eq:90}, it is precisely proportional to the thickness of the shell.  Fig-\eqref{fig-7} displays the nature of energy inside the thin shell for different values of $a$.

\subsection{Entropy} \label{subsec:C}

Mazur and Mottola \cite{Mottola/2002} studied the inner region and came to the conclusion that the entropy density is zero, representing the dependability of the condensate phase. The entropy of a charged gravastar in the intermediate region is computed as \cite{Pradhan/2023}

\begin{equation}\label{eq:91}
     \mathcal{S}= \int^{d+\epsilon}_{d} 4 \pi r^2 s(r) \sqrt{e^{B(r)}} dr.
\end{equation}
The entropy density $s(r)$ can be defined as follows using the standard thermodynamic relation $T \times s = p+\rho$ for a relativistic fluid with zero chemical potential and at the local temperature $T(r)$:

\begin{equation}\label{eq:92}
    s(r)=\frac{\kappa^2 K_{B}^2  T(r)}{4 \pi \hbar^2 }= \kappa \left(\frac{K_B}{\hbar} \right) \sqrt{\frac{p(r)}{2 \pi}}, 
\end{equation}
where  $\kappa$ is a dimensionless parameter.\\
Here we followed the treatment and argument given by Mazur and Mottola \cite{Mottola/2002}. We note that from the expression of $T\times s$ it is clear that in the core $\rho=-p$ the entropy is zero so the entire contribution of entropy is coming from the shell, which has been assumed to be a single quantum state (like Bose-Einstein condensed ground state) which gives contribution to gravastar.
Now equation \eqref{eq:91} becomes

\begin{equation}\label{eq:93}
   \mathcal{S}=\int^{d+\epsilon}_{d} 2 \sqrt{2 \pi}  \kappa \left(\frac{K_B}{\hbar} \right) a r\sqrt{\frac{3(r-3 \chi)}{3 a r+b r^3-6 a \chi}} \, dr. 
\end{equation}

Equation \eqref{eq:93} can also be written as
\begin{equation} \label{eq:94}
    \mathcal{S}=2 \sqrt{2 \pi}  \kappa \left(\frac{K_B}{\hbar} \right) a r \mathcal{J},
\end{equation}
where
\begin{equation}\label{eq:95}
 \mathcal{J} = \int^{d+\epsilon}_{d} \mathcal{D}(r) \, dr,  
\end{equation}
and 
\begin{equation}\label{eq:96} 
 \mathcal{D}(r)=\sqrt{\frac{3(r-3 \chi)}{3 a r+b r^3-6 a \chi}} .  
\end{equation}
 Evaluating the integral given in equation \eqref{eq:95} poses a significant challenge at present, so to solve the above integral let us take $F(r)$ to be the primitive of $D(r)$. Then equation \eqref{eq:95} becomes by using the fundamental theorem of integral calculus:
\begin{equation}\label{eq:97}
    \mathcal{J}=[F(r)]^{d+\epsilon}_{d}=F(d+\epsilon)-F(d).
\end{equation}

Expanding $F(d+\epsilon)$ in the Taylor series around `$d$' while keeping the linear order of $\epsilon$ from equation \eqref{eq:97}, we obtain, from \eqref{eq:93}, that

\begin{equation}\label{eq:98}
\mathcal{S}= 2 \sqrt{2 \pi}  \kappa a \left(\frac{K_B}{\hbar} \right)  r \times d \times \epsilon \sqrt{\frac{3(r-3 \chi)}{3 a r+b r^3-6 a \chi}}.
\end{equation}

\begin{figure}[h]
    \includegraphics[scale=0.6]{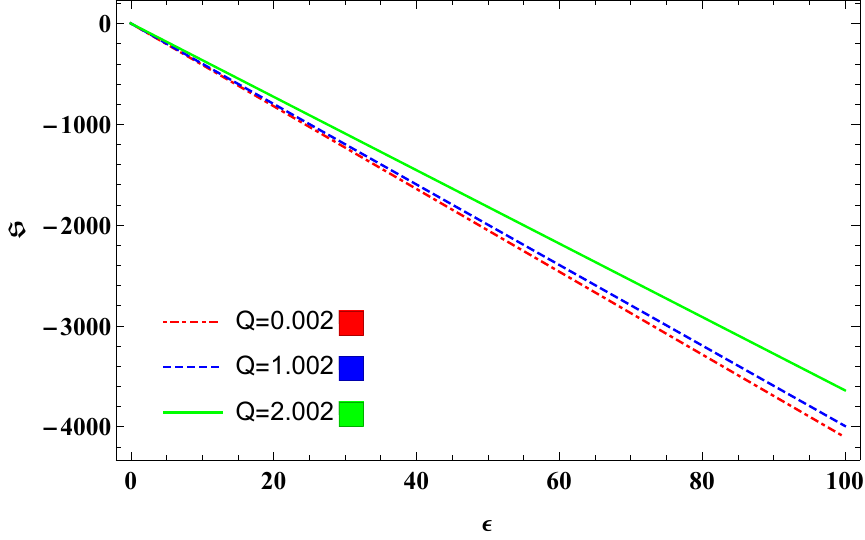}
    \caption{Variation of entropy inside the shell versus thickness of the shell of charged gravastar}
    \label{fig-8}
\end{figure}

As a result, we were successful in obtaining the entropy expression for the model we have suggested. One can infer from equation\eqref{eq:98} that if the thickness of the thin shell is $\epsilon<<d$, then $\mathcal{S}\approx \mathcal{O}(\epsilon)$. Fig-\eqref{fig-8} displays the variation of entropy with regard to the thin shell radius.

\subsection{The EoS parameter}\label{subsec:D}
The equation of state parameter $\omega$ can be written as \cite{Pradhan/2023} 
\begin{equation}
    \omega=\frac{P}{\varsigma}. \label{eq:99}
\end{equation}
Now by using equations \eqref{eq:73} and \eqref{eq:74}, we obtain
\begin{equation} \label{eq:100}
     \omega=\frac{1}{2}\left[\frac{\frac{-1+\frac{M}{\textbf{a}}}{\sqrt{-1 +\frac{2 M}{\textbf{a}}-\frac{Q^2}{\textbf{a}^2}}}-2 \textbf{\textbf{a}} \Tilde{\psi_0} }{ \left(\textbf{a} \Tilde{\psi_0}-\sqrt{-1+\frac{2 M}{\textbf{a}}-\frac{Q^2}{\textbf{a}^2}} \right)}\right].
\end{equation}

\begin{figure}
    \centering
    \includegraphics[scale=0.6]{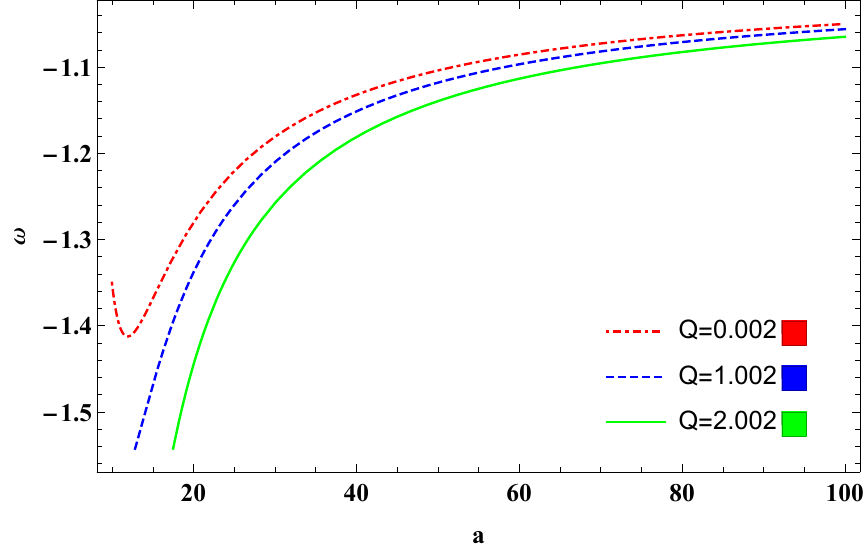}
    \caption{Variation of EoS $\omega$ with respect to $\textbf{a}$}
    \label{fig-9}
\end{figure}

The limitation $\frac{2M}{\textbf{a}}-\frac{Q^2}{\textbf{a}^2}< 1$, which is already satisfied by equation \eqref{eq:60}, is necessary to maintain the realness of $\omega$. Now, depending on the signature of the numerator or denominator of equation\eqref{eq:100}, $\omega$ may be positive or negative. Fig. \eqref{fig-9} displays the profiles of $\omega$ against `$\textbf{a}$'.

\section{ADM mass} \label{sec:IX}

We note that mass in general relativity can be a tricky concept as the gravitation field itself interacts with itself making it subtle to define how to define for general context. However, after the ADM formulation \cite{adm} or the Hamiltonian formulation of general relativity came it became somewhat clear that if we take the Hamiltonian and integrate it through space like Cauchy hypersurface we might be able to define a mass in such a way that incorporates the gravitational interactions. Note that an asymptotic flatness condition is needed for such a thing to happen and as we can see at $r\rightarrow \infty$ our Reissner-Nordstrom solution does indeed go to Minkowski metric, so we can be assured that we can indeed do the ADM mass formula. We note that the ADM formulation in $f(Q)$ gravity has already been done in \cite{Hu/2022} and authors have shown the equivalency of $f(Q)$ gravity and ordinary general relativity.\\
We have calculated the ADM mass for each of the two components of the gravastar which is an interior shell and outer region (as a thin shell in principle has a negligible thickness in this case so we cannot choose a space like Cauchy surface for $3+1$ splitting). The reason to consider the ADM mass for gravastar is the fact that we want to distinguish it from the black hole. As we will show the ADM mass for the charged gravastar is indeed different from the charged black hole due to its interior being de Sitter space.\\
We note that the formula for the ADM mass is given by \cite{Wald/1984},

\begin{equation}\label{eq:101}
M_{ADM}=\frac{1}{16\pi }\lim \limits _{r \rightarrow \infty }\sum _{\mu ,\nu =1}^3\int _{S}(\partial _{\mu }h_{\mu \nu }-\partial _{\nu }h_{\mu \mu })N^{\nu }dS .
\end{equation}
where $h_{\mu\nu}$ is the three-dimensional metric from the space like hyper-surface at $t$ being some constant time. $S$ is the two-sphere on which the integral has been performed and $N$ is the outward normal.\\
Now our metric is given in equation \eqref{eq:60} is in a  4-dimensional space-time manifold, over a constant time slice $\Sigma$, the embedded metric takes the form
    \begin{multline} \label{eq:102}
       h= ds^2_{\Sigma}=\left(1-\frac{2M}{r}+\frac{Q^2}{r^2}\right)^{-1}dr^2+r^2d\theta^2 \\
       +r^2\text{sin}^2\theta d\Phi^2.
    \end{multline}
Now, in general, the integral in the equation \eqref{eq:101} is quite difficult, but for spherical symmetric metric, it is quite easy as it is done in \cite{adm2}.\\
The calculation is pretty involved as one has to switch from polar coordinates to Cartesian coordinates and use the symmetry property to get the answer. Here we mention the final solution following the prescription given in \cite{adm2},
 for a space-time metric is given by,
\begin{equation} \label{eq:103}
    g=\varphi dr^2 +\chi(r)r^2d\Omega^2.
\end{equation}
Also given that $\varphi$ and $\chi$ reach the asymptotic flat space-time limit as $\varphi-1=o(r^{-\frac{1}{2}})$, $\chi-1=o(r^{-\frac{1}{2}})$ and $\partial_r\varphi=o(r^{-\frac{3}{2}})$, $\partial_r\chi=o(r^{-\frac{3}{2}}).$
 Then $M_{ADM}$ is given by, 
 \begin{equation}\label{eq:104}
    M_{ADM}= \lim\limits_{r \to \infty}\frac{1}{2}(-r^2\chi'+r(\varphi-\chi)).
 \end{equation}
For our case in outside, we take the Reissner-Nordstrom  solution so we get $\varphi(r)=\frac{1}{\left(1-\frac{2M}{r}+\frac{Q^2}{r^2}\right)}$ and $\chi(r)=1$, so calculating the limit we get,
\begin{equation}\label{eq:105}
\begin{gathered}
     M_{ADM}=\lim\limits_{r \to \infty}\frac{1}{2}\left[-r^2\chi'+r(\varphi-\chi)\right]\\
     =\lim\limits_{r \to \infty}\frac{1}{2}r\left[\frac{1}{\left(1-\frac{2M}{r}+\frac{Q^2}{r^2}\right)}-1\right]\\
     =\lim\limits_{r \to \infty}r\left[\frac{\frac{2M}{r}-\frac{Q^2}{r^2}}{2}\right] \\ = M-\frac{Q^2}{2r}
     \end{gathered}
\end{equation}
So, we get the contribution for ADM mass due to the outer region of the thin shell is given by,
\begin{equation}\label{eq:106}
    M_{ADM}=M-\frac{Q^2}{2r}.
\end{equation}
We know that to coincide with the Schwarzschild solution, $Q=0$ as it should happen. Also, we note that the presence of charge $Q$ can be seen in the formula for $M_{ADM}$.\\
Similarly, ADM mass in the interior region can be found by a similar formula as we note that  from equation \eqref{eq:83} we get,

    $$e^{B(r)}=\left( -\frac{Q^2}{r_1^3}+\frac{r^2}{2r_1^2}+\frac{br^2}{6\textbf{a}}-\frac{3 r^2 r_2}{2r_1^3}+\frac{2M r^2}{r_1^3}-\frac{b r^2 r_2^3}{6\textbf{a} r_1^3}\right)^{-1}.$$
   
So we can see the radial component of the metric is given in the form $e^{b(r)}=C(1+D r^2)^{-1}$, where $C$ and $D$ are functions made up of the constants $r_1$, $r_2$, $\textbf{a}$, $Q$ etc., as given in the above equation.\\
For inside the shell if we do the ADM mass calculation by using equation \eqref{eq:104} $\lim\limits_{r \to \infty}\frac{1}{2}r\left[\frac{C}{\left(1+Dr^2\right)}-1\right]$.\\
We can see that the limit diverges for the arbitrary values of $D$, which is not at all unexpected as the Schwarzschild-de Sitter metric's radial coordinate has the form $\frac{1}{1-\frac{2M}{r}-\frac{1}{3}\Lambda r^2}$, here also similarly the ADM mass diverse as this is not asymptotically flat.\\
This shows that for the interior region, ADM mass does not make sense as the metric is not asymptotically flat, as we have also taken de Sitter space-time inside.

\section{Discussion and Conclusion} \label{sec:X}

In this article, we have studied the analytical solution of charged gravastar in $f(\mathcal{Q})$ gravity. As we have mentioned motivated by the black hole ``no-hair" theorem we have taken charged gravastar as we are modelling gravastar as an alternative to the black hole after a gravitational collapse. We have started with the three regions of gravastar which are interior (EoS=-1), shell (EoS=1), and exterior regions ( Reissner-Nordstrom metric as there is a non-zero charge). Using two junction conditions we have found the analytic solution to the metric coefficient which is drawn in fig-\eqref{fig-1}  and fig-\eqref{fig-2}. We also note that as the other region is Reissner-Nordstrom space-time so it is asymptotically flat (i.e. at $r\rightarrow \infty$ the metric becomes Minkowski), one needs this to find the ADM mass of a system. As our solutions come with arbitrary integration constants, and we are keeping the discussion for generally charged gravastar in table-I we have taken some typical values of stellar mass black holes for bookkeeping.\\
In the next part, we have taken the Israel junction condition for the thin shell (neglecting the thickness) and found out the energy density and pressure corresponding to the thin shell which we have plotted in fig-\eqref{fig-3} and fig- \eqref{fig-4}  respectively. We have also calculated the potential across the thin shell and we note that it can be used as a phenomenological probe to detect not just gravastar but to distinguish between black hole and gravastar. We can see from the fig-\eqref{fig-5} that the potential does have a varying $r_{min}$ with charge and after a certain critical charge, the min of $r_{min}$ vanishes so this denotes the ISCO (innermost stable circular orbit). It is widely known that ISCO has been a very successful tool for finding black hole accretion disks and rotation etc. \cite{Narayan/1994}. So one can in principle distinguish between gravastar and black hole by shifting of Fe (iron) line and measuring the ISCO which is different from the  Reissner-Nordstrom black hole.\\
We also provide another phenomenological observation that in principle can distinguish between black holes and gravastars. We have calculated the angle of deflection for the gravastar interior region and exterior region, even though the exterior region equation \eqref{eq:86} $I(r_0)$ is the same as that of the  Reissner-Nordstrom black hole, but we can see 
equation \eqref{eq:84} $I(r_0)_{inside}$ is much different from that of the  Reissner-Nordstrom black hole. In simple words, as gravastar does not have any event horizon it is expected that if certain photons can penetrate the shell (which it can as stiff matter does not violate any causality), we can indeed use a next-generation event horizon telescope to detect such photons in this way we can distinguish between the gravastar and black hole.\\
Finally, in order to investigate the properties of the gravastar we have used the standard methods developed by \cite{Mazur/2004} to get an overall understanding of the properties of the gravastar. In fig-\eqref{fig-6}the variation of proper length vs thickness ($\epsilon$) is given as we can see it's monotonically increasing as expected. In fig-\eqref{fig-7}  and fig-\eqref{fig-8} we have plotted the functional form of energy and entropy inside the shell as a function for thickness ($\epsilon$). We have noted in all three cases that the proper length, energy density, and entropy have exactly similar features as given by \cite{Mazur/2004}. So we can safely conclude that our solution of charge gravastar in $f(\mathcal{Q})$ gravity does indeed share similar characteristics of a gravastar \cite{Mazur/2004}.\\
In fig-\eqref{fig-9} we have shown the variation of EoS as a function of $\textbf{a}$. We see that for large $\textbf{a}$ it is $-1$ which is expected from the fact we have started with the de Sitter metric inside. We can also see that as $\textbf{a}$ goes smaller the EoS state becomes less than $-1$, that is it goes to the phantom region. Even though in the cosmological context there are problems with dealing with phantom regions, in this case, it just gets to show that to stabilize the gravastar we need more than just ``Dark Energy".\\
 We have also calculated the ADM mass sec \ref{sec:IX}  for the gravastar inside and outside of the shell. We note that the ADM formulation requires $3+1$ splitting of the space-time, as by the assumption the thin shell is really "thin" to contribute to a $3-$dimensional Cauchy hypersurface. We have noted that outside the ADM mass is similar to the  Reissner-Nordstrom black hole as expected, but inside it diverges it has to do with the fact inside metric is de Sitter so it is not asymptotically flat so can not expect the ADM mass to be finite there.\\
 As far as the observational aspect of the gravastar comes it is a very active research field where the last words have been said. First, we note that after detecting the gravitational waveform merging black hole binaries in 2016 \cite{Abbott/2016}, we have officially entered the so-called ``multi-messenger era " of astronomy, where we can observe our universe using probes very different (like a neutrino, gravitational wave, etc.) than EM waves (radio telescope, X-ray telescope, infrared telescope, etc. just to name a few). Definitive observational proof of gravitational waves indeed gives hope that maybe one can distinguish gravastar from black holes via GW observation. In the work by Cardoso et al. \cite{Cardoso/2016}  it has been shown that to study the event horizon and its quantum mechanical phenomenon one can use late time ringdown of gravitational wave. The authors also conclude how the study of late time ringdown can be used to rule out the alternatives of the black holes like gravastars, bose stars, etc. The work by Chirenti et al. \cite{Chirenti/2016}  takes it one step further by showing it is very difficult to see the difference between the black hole and gravastar as just like the black hole for gravastar we do not have any results about the perturbations with respect to some fixed background. The authors have studied the data from GW150914 to give a heuristic bound on the confidence by which we can say whether it was a black hole merger or gravastar.\\
 Apart from gravitational waves, there are other proposals to distinguish gravastars from black holes as Broderick \cite{Broderick/2007} and Narayan \cite{Narayan/1994} have proposed that one can do a fallback disk analysis of gravastars just like the acceleration disk analysis of black hole and the authors also give a bound on the observational constraints to distinguish gravastars from black holes. We also note that Uchikata et al. \cite{Uchikata/2016}  have suggested studying the observational properties of gravastar from an astrophysical object point of view. As we have noted earlier gravastar do not have an event horizon so it is natural to calculate its tidal deformability and I-Love-Q relations for a small perturbation like a neutron star. The authors have done the calculation for love number as well as I-Love-Q relation of gravastar under small perturbations and have given us a phenomenological bound on under what conditions one can distinguish the gravastar and block hole via observation.\\
   We note that even though this is an exciting era we can ask whether the gravstars can be distinguished from the black holes by multi-messenger astronomy. In our work, we propose that one can indeed distinguish them if some low-energy photons from the interior regions of gravastars can penetrate the shell and come outside. We can use the angle of deflection formula to see the difference from the black hole. \\
   We note that after the black hole image formed by the Event Horizon Telescope \cite{Akiyama/2019}  it is natural to ask whether such a shadow can be found for gravastar or not. In our article we have found the deflection angle for the interior region and hope that such a phenomenological study will help next-generation radio telescopes to search for more abnormality from black hole shadow and in future can indeed distinguish between gravastar and black hole. \\
Finally, we note that also one can test the viability of the various modified gravity from the shadow of black hole \cite{Mizuno/2018}, so maybe we can hope future generation radio telescopes will not just give the shadow of gravastar but will also provide whether the modification of Einstein gravity by $f(\mathcal{Q})$ gravity is indeed right choice or not.

\textbf{Data availability} There are no new data associated with this article.

\acknowledgments DM expresses gratitude to the BITS-Pilani, Hyderabad campus, India, for the financial support. SG acknowledges Council of Scientific and Industrial Research (CSIR), Government of India, New Delhi, for junior research fellowship (File no.09/1026(13105)/2022-EMR-I). PKS  acknowledges the National Board for Higher Mathematics (NBHM) under the Department of Atomic Energy (DAE), Govt. of India for financial support to carry out the Research project No.: 02011/3/2022 NBHM(R.P.)/R \& D II/2152 Dt.14.02.2022 and IUCAA, Pune, India for providing support through the visiting Associateship program. We are very much grateful to the honorable referees and to the editor for the illuminating suggestions that have significantly improved our work in terms of research quality, and presentation.

\end{document}